\documentstyle[general_cite,psfig]{mn}
\title{X-ray luminosities of galaxies in groups}
\author[S. F. Helsdon et al.]
       {Stephen F. Helsdon\thanks{E-mail: sfh@star.sr.bham.ac.uk}$^1$, Trevor
         J. Ponman$^1$, Ewan O'Sullivan$^1$ \cr 
         and Duncan A. Forbes$^{1,2}$\\
  $^1$School of Physics and Astronomy, University of
        Birmingham, Edgbaston, Birmingham B15 2TT, UK\\
  $^2$ Astrophysics \& Supercomputing, Swinburne University of Technology,
        Hawthorn VIC 3122, Australia\\}
 \date{Accepted 2000 ??.
      Received 2000 ??;
      in original form 2000 ??}

\pagerange{\pageref{firstpage}--\pageref{lastpage}}
\pubyear{2000}
\def\etal{{\it et al. }}

\begin{document}

\maketitle

\label{firstpage}

\begin{abstract}
  We have derived the X-ray luminosities of a sample of galaxies in groups,
  making careful allowance for contaminating intragroup emission. The
  L$_X$:L$_B$ and L$_X$:L$_{FIR}$ relations of spiral galaxies in groups
  appear to be indistinguishable from those in other environments, however
  the elliptical galaxies fall into two distinct classes. The first class
  is central-dominant group galaxies which are very X-ray luminous, and may
  be the focus of group cooling flows.  All other early-type galaxies in
  groups belong to the second class, which populates an almost constant
  band of L$_X$/L$_B$ over the range 9.8 $< \log L_B <$ 11.3. The X-ray
  emission from these galaxies can be explained by a superposition of
  discrete galactic X-ray sources together with a contribution from hot gas
  lost by stars, which varies a great deal from galaxy to galaxy.  In the
  region where the optical luminosity of the non-central group galaxies
  overlaps with the dominant galaxies, the dominant galaxies are over an
  order of magnitude more luminous in X-rays.
  
  We also compared these group galaxies with a sample of isolated
  early-type galaxies, and used previously published work to derive
  L$_X$:L$_B$ relations as a function of environment. The non-dominant
  group galaxies have mean L$_X$/L$_B$ ratios very similar to that of isolated
  galaxies, and we see no significant correlation between L$_X$/L$_B$ and
  environment. We suggest that previous findings of a steep L$_X$:L$_B$
  relation for early-type galaxies result largely from the inclusion of
  group-dominant galaxies in samples.
\end{abstract}


\begin{keywords}
galaxies: ISM -- galaxies: elliptical and lenticular -- galaxies: spiral -- 
X-rays: galaxies
\end{keywords}


\section{Introduction}
\label{sec:intro}

Many galaxies in the Universe are found in galaxy groups (e.g.
\pcite{tully87}). These group galaxies show some differences from their
counterparts in the field. For example, there is evidence that some spirals
interact with their environment \cite{verdesmontenogro99}, and with each
other (e.g.  \pcite{rubin91,forbes92}). Early-type galaxies in groups are
less likely to have boxy isophotes and more likely to have irregular
isophotes than comparable ellipticals in other environments \cite{zepf93w}.
Many of these features are likely to be due to the effects of the group
environment in which the galaxy is found, and it is possible that the X-ray
properties of galaxies could be affected by the group environment as well.

There are two main sources for the X-ray emission from normal galaxies:
stellar sources and hot diffuse interstellar gas.  Emission from
bright early-type galaxies is dominated by the hot diffuse component (e.g.
\pcite{forman85,trinchieri85}) whilst in late-type and less luminous early
types, the emission is primarily from stellar sources (e.g.
\pcite{fabbiano85,kim92}). There have been a number of studies of the X-ray
properties of both spiral (e.g.  \pcite{fabbiano85,fabbiano88}) and
elliptical (e.g. \pcite{canizares87,eskridge95,brown98,beuing99}) galaxies,
which show that while the X-ray luminosity of late-type galaxies scales
roughly with the optical luminosity, the same relation for early-type
galaxies is considerably steeper and shows much more scatter.

If environment does have a significant effect on galaxy X-ray properties,
this could explain some of the scatter observed in the L$_X$:L$_B$
relations for galaxies. A galaxy moving through the intragroup medium may
undergo ram pressure stripping \cite{gunn72} or viscous stripping
\cite{nulsen82}, thus decreasing the X-ray luminosity without much
affecting the stellar luminosity. Conversely, material cooling onto a
galaxy from the intragroup medium could enhance the observed galaxy
X-ray luminosity \cite{canizares83}.  The effects of galaxy winds may also
play an important role.  \scite{brown00} found a positive correlation
between L$_X$/L$_B$ and local galaxy volume density, for early-type
galaxies using {\it ROSAT} data. They suggested that the X-ray luminosity
of early-type galaxies is enhanced in higher density environments. This is
in contrast to the earlier work of \scite{white91} who claimed detection
of the opposite effect, using {\it Einstein} data.

In practice, study of the X-ray properties of galaxies within groups is
complicated by the presence of X-ray emission from a hot intragroup medium
in many bound groups \cite{mulchaey96,ponman96a,helsdon00}. Many previous
studies of the X-ray emission from galaxies, especially early-type
galaxies, which predominate in X-ray bright groups, have failed to allow
for this problem. This can result in poor estimates for the luminosity,
extent and spectral properties of galaxy emission. In the present paper we
attempt to remedy this by carefully removing group emission, which in some
cases involves using multicomponent models of the surface brightness
distribution \cite{helsdon00} within a group. By comparing the properties
of these galaxies to those in low density environments we seek to gain some
insight into the processes operating in the group environment.  Throughout
this paper we use H$_0$=50 km s$^{-1}$ Mpc$^{-1}$, and all errors are
1~$\sigma$.


\section{A sample of group galaxies}
\label{sec:sample}

In order to investigate the X-ray properties of galaxies in groups, it is
first necessary to define a sample of groups with available X-ray data. For
this study we have combined the sample of groups studied by
\scite{helsdon00} with additional data for compact galaxy groups.
We use only galaxies from groups in which a hot intragroup medium
is detected, since this confirms that these groups are genuine mass
concentrations, as opposed to chance line-of-sight superpositions.

\scite{helsdon00} compiled a sample of 24 X-ray bright groups observed with
the ROSAT PSPC. In the case of compact galaxy groups, the tight
configuration of the major group galaxies (typically separated by only a
few arcminutes) can lead to serious problems of confusion, and
contamination of galaxy fluxes by diffuse group emission. We therefore have
added to the sample all the Hickson Compact Group (HCG; \pcite{hickson82})
galaxies observed by the ROSAT HRI. The HCGs thus added were 15, 16, 31,
44, 48, 51, 62, 68, 90, 91, 92, and 97. In the case of overlap between the
HRI data and the PSPC data of \scite{helsdon00} (HCGs 62, 68 and 90), the
HRI data were preferentially used, except in the case of HCG 62. In the
case of HCG 62, the diffuse group emission is so bright that no galaxies
could be detected, and upper limits were dominated by uncertain systematic
errors in modeling the diffuse flux.  However the PSPC data for this system
clearly show a central component, fairly distinct from the surrounding
group emission; a structure very similar to that of a number of other PSPC
groups. We have therefore omitted this group from the HRI dataset and used
the PSPC data instead. This resulted in a final sample of 33 groups, 11
HCGs with HRI data and 22 other groups with PSPC data. Note that the PSPC
data contains two compact groups (HCG 42 \& HCG 62), which are listed under
the names given in \scite{helsdon00} (NGC 3091 \& NGC 4761).

Principal galaxies in the HCGs are listed with types and magnitudes in
\scite{hickson89}. The sample for which we were able to derive useful X-ray
flux estimates or upper limits, contains 46 galaxies in 11 HCGs. For the
loose groups, we wished to include only galaxies residing within the denser
inner regions, where environmental effects might be expected. We therefore
searched the NASA/IPAC Extragalactic Database (NED) for galaxies lying
within a third of the group virial radius in projection on the sky, and
having recession velocities within three times the group velocity
dispersion (3$\sigma_g$) of the catalogued group mean. The virial radius of
each group (typically $\sim$ 1.1 Mpc) is calculated as described in
\scite{helsdon00}, which also lists $\sigma_g$ for each group. All galaxies
with a listed optical magnitude and galaxy type were initially included. A
number of extra galaxies were included with galaxy types from
\scite{zabludoff98}, who have carried out multifibre spectroscopy on 9 of
the groups. Also included were a few galaxies whose galaxy types were
easily determined from Digitized Sky Survey images. This gave a total of
114 galaxies (in 22 loose groups ) for which we were able to derive useful
X-ray flux estimates or upper limits.  These were split into four broad
morphological categories i.e. spiral, elliptical, lenticular and irregular.
Adding the HCG galaxies gives a total sample of 160 galaxies in 33 groups.
This sample should not be regarded as being statistically complete in any
way, but rather a reasonably representative sample of galaxies in collapsed
groups.


\section{Data reduction}
\label{sec:red}

\subsection{ROSAT HRI data}
\label{sec:red:com}

HRI data were reduced using standard \textsc{asterix} software, and
binned into 3\arcsec pixel images, using the full energy range of the
detector. Due to the compact nature of HCGs, almost all galaxies studied
lie within 5\arcmin of the pointing axis, so that vignetting affects their
detected fluxes by $<3$\%, and no correction was applied for this.
                                  
The high resolution of the HRI generally provides a clear separation
between the galaxy and group emission, and in many cases the latter was
undetectable. Radial profiles centred on each galaxy were used to determine
the radial extent of galaxy-related emission, and the galaxy count rate was
then extracted from a circle encompassing this emission. In most cases the
background was flat, and was determined from a large source-free region
near the centre of the field.  A few systems with bright diffuse X-ray
emission were more difficult. Wherever diffuse group emission might
contribute more than 5\% contamination to the galaxy fluxes (i.e. HCG 51,
90 and 97), we fitted an elliptical model to the diffuse flux distribution,
and this was then used as a background model when deriving galaxy count
rates or upper limits. In the case of HCG 92, the diffuse emission is very
irregular, since it apparently arises from intergalactic shocks, rather
than a hydrostatic intragroup medium \cite{pietsch97}. Since we were unable
to reliably model this emission, three of the five galaxies in this system
had to be excluded from our sample.

Source count rates, together with Poisson errors were extracted and
background subtracted. Cases where the source counts exceeded twice the
Poisson error on the background ($\sigma_b$) were deemed to be detections.
For non-detected sources, an upper limit of $3\sigma_b$ is returned,
normally calculated within a circle of radius 18\arcsec.  All count rates
were converted to unabsorbed bolometric luminosities assuming a 1~keV
Raymond \& Smith (1977) model with 0.25 solar metallicity, and distances
given in Table~1. This model was selected as it should provide a reasonable
description of the spectral properties of an early-type galaxy
\cite{davis96b}. For the spiral and irregular galaxies a hotter spectral
model may be more appropriate, but for simplicity the same spectral model
was used for all galaxies. If a hotter temperature of 5~keV is used
instead, the luminosities of the late-type galaxies would rise by about
0.24 dex. B-band luminosities were derived using magnitudes from NED, and
assuming a solar blue luminosity of 5.41$\times 10^{32}$ erg s$^{-1}$.

\subsection{ROSAT PSPC data}
\label{sec:red:loo}

Previous work on the X-ray properties of galaxies tends to use
one of two approaches: (a) trace the extent of the emission from the
galaxy position out to a background level, and include all this
emission as the galaxy emission (e.g. \pcite{beuing99}), or (b)
extract the emission within some standard radius, such as four times
the optical effective radius of the galaxy (e.g.  \pcite{brown98}).
The first method may significantly overestimate the X-ray luminosity
of any galaxy located near the centre of a group, while the second
takes no account of the X-ray surface brightness profile of the
galaxy, and so could result in an over- or underestimate of the true
luminosity.

For the galaxies in loose groups, we adopted one of two different methods
to derive X-ray luminosities, depending on their position within the group,
as described in sections \ref{cenred} and \ref{otherred} below.
2-dimensional models of the group emission were available for each of the
22 loose groups \cite{helsdon00}. For a number of systems, these model fits
indicated the presence of a central cusp, in addition to more extended
emission associated with the group as a whole. Such central components
invariably coincide with a central galaxy. For these cases we take the
central component to be the emission associated with the galaxy. For these
and all the other galaxies we are able to use the group models to remove
the group contribution to the galaxy flux. In the case of HCG 62, the
central X-ray component is centred on the brightest galaxy, HCG 62a, but
also encompasses a second galaxy (HCG 62b). We identify this X-ray
component with HCG 62a, but inclusion of the second galaxy would only
reduce the derived L$_X$/L$_B$ by 50\% (-0.18 in log(L$_X$/L$_B$)).

For the remainder of this paper, galaxies coincident with clear central
X-ray cusps will be referred to as central-dominant galaxies. The
brightest group galaxy (BGG) refers to the optically brightest member. It
should be noted that a central-dominant group galaxy will generally be a
BGG, although a BGG will not necessarily be a central-dominant galaxy
as defined here.

\subsubsection{Central-dominant galaxies\label{cenred}}

For 11 of the groups, \scite{helsdon00} have derived two-component fits,
involving an extended group component and a central component coincident
with a central, optically bright galaxy. In a few of these cases the
central component was poorly resolved by the {\it ROSAT} PSPC so we
examined HRI data if available, in order to check the properties of the
central components. These checks showed that use of the HRI data did not
significantly alter the properties of the central components derived by the
PSPC. Thus for these 11 cases we have used the central component from the
PSPC fits to derive the luminosity associated with the central galaxy. We
have also added to these 11, the group NGC 4325, for which we have fitted a
2 component model. \scite{helsdon00} found that a one component model was
adequate for this system, however the addition of a second component does
significantly improve the fit ($\Delta$Cash~=~43.6). The luminosity for
each of these 12 cases was derived by using the 2D models to calculate the
fraction of the total luminosity of the group emission contained in the
central component. This contribution was then extracted from the group
total luminosity given in \scite{helsdon00} to give the luminosity of the
galaxy itself.

\subsubsection{Other group galaxies\label{otherred}}

Initial reduction of the ROSAT PSPC data was carried out as in
\scite{helsdon00}. An image of the group was then generated and a radial
profile produced for each galaxy listed in Table~1.  These profiles were
examined individually and used to identify the radius to which emission
could be traced. In cases where poor statistics prevented extraction of a
reliable profile, a fixed radius of 20 kpc was used.  Examination of both
the image and radial profiles also enabled the identification of cases
where the X-ray data may be contaminated by a nearby bright source,
possible cases of source overlap or cases where the galaxies may be
partially obscured by the PSPC support ring. Such cases (20 galaxies in
total) were excluded from further analysis. Of these 20 galaxies, 8 were
removed due to the PSPC ring and the majority of the remainder were
galaxies probably confused with a central galaxy X-ray component. In almost
all these confused cases the central galaxy was much brighter than the
removed galaxy. Even if all the optical light from these other galaxies was
included, the derived L$_X$/L$_B$ of the affected central galaxy would
typically only change by 5\% (0.02 in log space. HCG 62a, discussed
earlier, is the case with the largest change).

For each galaxy, a count rate was extracted within the radius as determined
above, and then corrected to an on-axis count rate. As all these galaxies
lie within X-ray bright groups, a fraction of the count rate will be due to
group emission. To correct for this, the best fitting 2D surface brightness
profile models derived by \scite{helsdon00} were used to determine the
contribution from the group, which was then subtracted from the count rate
extracted at the position of the galaxy.

Detections and upper limits were derived, and converted to source
luminosities (bolometric), in the way described above for compact group
galaxies.  The luminosities of all the group galaxies in our sample are
listed in Table~1.

\begin{table*} 
\begin{tabular}{p{3.2cm}cllccp{3.2cm}cll}
Galaxy name & Type & L$_B$ & L$_X$ & & & Galaxy name & Type & L$_B$ & L$_X$ \\
\hline
\hline                              
\multicolumn{3}{l}{{\bf~NGC 315 group} 96 Mpc}                        & & & & \multicolumn{3}{l}{{\bf~NGC 4065 group} 151 Mpc}               &                       \\
NGC 315                 &   E   &  11.27        &   $<$41.09            & & & NGC 4065                &   E   &  11.11       &   41.44 $\pm$ 0.30  \\
NGC 311                 &   S0  &  10.55        &   40.46 $\pm$ 0.22    & & & NGC 4061                &   E   &  10.90       &   41.28 $\pm$ 0.12  \\
\multicolumn{3}{l}{{\bf~NGC 383 group} 102 Mpc}               &                       & & & NGC 4072                &   S0  &  10.30       &   $<$40.88           \\
NGC 383                 &   S0  &  10.95        &    42.30 $\pm$ 0.06   & & & NGC 4060                &   S0  &  10.30       &   $<$40.82           \\
NGC 386                 &   E   &  10.07        &   40.29 $\pm$ 0.28    & & & NGC 4066                &   E   &  10.96       &   41.94 $\pm$ .07   \\
NGC 380                 &   E   &  10.76        &   41.58 $\pm$ .04     & & & PGC 038163              &   S   &  10.18       &   $<$40.84           \\
NGC 385                 &   S0  &  10.63        &   41.16 $\pm$ .08     & & & NGC 4074                &   S0  &  10.38       &   41.37 $\pm$ 0.14  \\
UGC679                  &   S   &  9.56         &   $<$40.440           & & & NGC 4076                &   S   &  10.80       &   $<$40.80           \\
NGC 375                 &   E   &  9.91         &   40.77 $\pm$ 0.18    & & & NGC 4070                &   E   &  10.89       &   40.96 $\pm$ 0.31  \\
NGC 384                 &   E   &  10.58        &   41.20 $\pm$ .07     & & & \multicolumn{3}{l}{{\bf~NGC 4073 group} 136 Mpc}               &                       \\
NGC 388                 &   E   &  10.03        &   $<$40.42            & & & NGC 4073                &   E   &  11.49        & 43.11 $\pm$ 0.02      \\
NGC 379                 &   S0  &  10.63        &   41.20 $\pm$ .07     & & & NGC 4139                &   S0  &  10.58       &   $<$40.74           \\
PCC S34-111:LLB96 402   &   E   &  9.56         &   $<$40.34            & & & NGC 4063                &   S0  &  10.45       &   41.27 $\pm$ 0.18  \\
PCC S34-111:LLB96 265   &   S0  &  10.29        &   $<$40.23            & & & NGC 4077                &   S0  &  10.83       &   41.50 $\pm$ 0.12  \\
\multicolumn{3}{l}{{\bf~NGC 524 group} 50 Mpc}               &                       & & & PGC 038154              &   S0  &  10.48       &   $<$40.77           \\
NGC 524                 &   S0  &  11.06        &   $<$40.50            & & & \multicolumn{3}{l}{{\bf~NGC 4261 group} 53 Mpc}               &                       \\
NGC 518                 &   S   &  9.92         &   $<$40.08            & & & NGC 4261                &   E   & 11.09         &  41.74 $\pm$ 0.04     \\
NGC 532                 &   S   &  9.99         &   $<$40.09            & & & NGC 4264                &   S0  &  10.15       &   40.47 $\pm$ 0.13  \\
\multicolumn{3}{l}{{\bf~NGC 533 group} 107 Mpc}               &                       & & & PGC 039655              &   i   &  9.49        &   $<$40.18           \\
NGC 533                 &   E   &  11.35        &    42.61 $\pm$ 0.03   & & & NGC 4257                &   S   &  9.68        &   $<$40.17           \\
NGC 0533:ZM98 0034      &   E   &  9.62         &   $<$40.88            & & & PGC 039639              &   I   &  8.91        &   $<$40.17           \\
NGC 0533:ZM98 0046      &   E   &  9.61         &   $<$40.54            & & & PGC 039708              &   S0  &  9.64        &   40.22 $\pm$ 0.18  \\
NGC 0533:ZM98 0027      &   E   &  9.55         &   $<$40.81            & & & NGC 4269                &   S0  &  10.06       &   40.29 $\pm$ 0.18  \\
NGC 0533:ZM98 0017      &   S   &  9.84         &   $<$40.50            & & & NGC 4260                &   S   &  10.55       &   40.56 $\pm$ 0.13  \\
NGC 0533:ZM98 0026      &   E   &  9.59         &   $<$40.48            & & & VCC 0405                &   E   &  7.63        &   $<$40.16           \\
\multicolumn{3}{l}{{\bf~NGC 741 group} 106 Mpc}               &                       & & & VCC 0315                &   S   &  9.64        &   $<$40.20           \\
NGC 741                 &   S0  &  11.4         &   41.90 $\pm$ 0.07    & & & NGC 4266                &   S   &  9.80        &   $<$40.25           \\
NGC 0741:ZM98 0009      &   E   &  10.19        &   $<$40.82            & & & PGC 039532              &   I   &  9.23        &   $<$40.27           \\
NGC 0741:ZM98 0023      &   E   &  9.43         &   $<$40.78            & & & PGC 039711              &   I   &  8.75        &   $<$40.31           \\
NGC 0741:ZM98 0010      &   E   &  10.07        &   $<$40.98            & & & NGC 4287                &   S   &  9.77        &   $<$39.83           \\
NGC 0741:ZM98 0014      &   S   &  10.03        &   $<$40.53            & & & \multicolumn{3}{l}{{\bf~NGC 4325 group} 163 Mpc}               &                       \\
NGC 0741:ZM98 0011      &   E   &  10.07        &   40.62 $\pm$ 0.30    & & & NGC 4320                &   S   &  10.70       &   $<$41.33           \\
\multicolumn{3}{l}{{\bf~NGC 1587 group} 77 Mpc}               &                       & & & NGC 4325:ZM98 0030      &   E   &  10.09       &   $<$41.06           \\
NGC 1587                &   E   &  10.88        &   $<$40.79            & & & NGC 4325                &   E   &  10.89        &  42.88 $\pm$ 0.06     \\
NGC 1588                &   E   &  10.40        &   $<$40.96            & & & \multicolumn{3}{l}{{\bf~NGC 4636 group} 33 Mpc}               &                       \\
NGC 1589                &   S   &  10.84        &   $<$40.80            & & & NGC 4636                &   E   &  11.04        &   42.03 $\pm$ 0.02    \\
PGC 015369              &   I   &  8.76         &   $<$40.62            & & & \multicolumn{3}{l}{{\bf~NGC 4761 group (HCG 62)}} 105 Mpc     &                       \\
\multicolumn{3}{l}{{\bf~NGC 2563 group} 107 Mpc}               &                 & & & NGC 4761                &   S0  &  10.9        &   42.78 $\pm$ 0.02   \\
NGC 2562                &   S   &  10.69        &   40.84 $\pm$ 0.17    & & & PGC 043760              &   E   &  9.83        &   $<$40.63           \\              
NGC 2563                &   S0  & 11.04         &  41.7 $\pm$ 0.07      & & & HCG 062:ZM98 0036       &   E   &  9.70        &   $<$40.58           \\              
NGC 2563:ZM98 0016      &   S   &  10.10        &   $<$40.48            & & & HCG 062:ZM98 0022       &   S   &  9.99        &   41.01 $\pm$ 0.13  \\               
NGC 2563:ZM98 0033      &   S   &  9.96         &   $<$40.70            & & & \multicolumn{3}{l}{{\bf~NGC 5129 group} 150 Mpc}               &                       \\      
NGC 2560                &   S   &  10.52        &   40.60 $\pm$ 0.27    & & & NGC 5129:ZM98 0007      &   S   &  10.38       &   $<$41.16           \\              
PGC 023448              &   E   &  10.10        &   $<$40.47            & & & NGC 5129                &   E   &  11.33       &   41.31 $\pm$ 0.25  \\               
PGC 023391              &   S   &  10.32        &   $<$40.35            & & & \multicolumn{3}{l}{{\bf~NGC 5176 group} 150 Mpc}               &                       \\      
\multicolumn{3}{l}{{\bf~NGC 3091 group (HCG 42)} 91 Mpc}     &                & & & NGC 5176                &   S0  &  10.38       &   $<$41.44           \\              
NGC 3091                &   E   & 11.31         &  42.04 $\pm$ 0.04     & & & NGC 5177                &   S0  &  10.38       &   40.90 $\pm$ 0.34  \\               
NGC 3096                &   S0  &  10.36        &   40.48 $\pm$ 0.23    & & & NGC 5171                &   S0  &  11.05       &   41.42 $\pm$ 0.30  \\               
HCG 042:ZM98 0046       &   S   &  9.27         &   $<$40.38            & & & NGC 5179                &   S0  &  10.58       &   41.12 $\pm$ 0.19  \\               
\multicolumn{3}{l}{{\bf~NGC 3607 group} 33 Mpc}               &                 & & & NGC 5178                &   S0  &  10.63       &   $<$41.41           \\              
NGC 3608                &   E   &  10.54        &   $<$40.24            & & & \multicolumn{3}{l}{{\bf~NGC 5846 group} 42 Mpc}               &                       \\      
NGC 3607                &   S0  &  10.90        &   40.12 $\pm$ 0.31    & & & NGC 5845                &   E   &  10.04       &   $<$40.35           \\              
NGC 3605                &   E   &  9.97         &   39.71 $\pm$ 0.32    & & & NGC 5850                &   S   &  10.82       &   40.57 $\pm$ 0.15  \\               
PGC 034419              &   S   &  9.42         &   $<$40.04            & & & NGC 5839                &   S0  &  9.96        &   $<$40.35           \\              
NGC 3599                &   S0  &  10.10        &   $<$40.17            & & & NGC 5846                &   E   & 11.07         & 41.51 $\pm$ 0.05      \\            
\multicolumn{3}{l}{{\bf~NGC 3665 group} 53 Mpc}               &                 & & & NGC 5846:ZM98 0015      &   S   &  9.20        &   $<$40.37           \\              
NGC 3665                &   S0  &  10.92        &   $<$40.61            & & & NGC 5848                &   S0  &  9.53        &   $<$40.45           \\              
NGC 3658                &   S0  &  10.38        &   40.52 $\pm$ 0.31    & & &                         &       &               &                       \\ 
\hline
\end{tabular}
\caption{\label{tab:data}The galaxy data. Names, types and B-band
  luminosities are derived from values given in the NED. L$_X$ units are
  bolometric log (erg s$^{-1}$), L$_B$ units are log (B-band Solar). Upper 
  limits are 3$\sigma$ above the background. Errors are 1 $\sigma$ and
  based on photon counting statistics.}
\end{table*}

\begin{table*}
{\bf Table~\ref{tab:data}} {\it continued}
\begin{tabular}{p{3.2cm}cllccp{3.2cm}cll}
Galaxy name & type & L$_B$ & L$_X$ & & & Galaxy name & type & L$_B$ & L$_X$ \\
\hline
\hline                    
\multicolumn{3}{l}{{\bf~NGC 6338 group} 171 Mpc}               &                       & & & \multicolumn{3}{l}{{\bf~HCG 48} 56 Mpc}               &                       \\
NGC 6338                &   S0  & 11.3          & 43.28 $\pm$ 0.02      & & & HCG 48a                 &   E   &  10.54       &   40.681 $\pm$ 0.106  \\
NGC 6345                &   S0  &  10.49        &   41.21 $\pm$ 0.28    & & & HCG 48b                 &   S   &  9.69        &   41.459 $\pm$ .021   \\
IC 1252                 &   S   &  10.37        &   $<$41.19            & & & HCG 48c                 &   S0  &  9.13        &   $<$40.079           \\
NGC 6346                &   E   &  11.16        &   $<$41.18            & & & HCG 48d                 &   E   &  8.69        &   $<$40.079           \\
NPM1G +57.0229          &   S   &  10.29        &   41.02 $\pm$ 0.30    & & & \multicolumn{3}{l}{{\bf~HCG 51} 155 Mpc}               &                       \\
\multicolumn{3}{l}{{\bf~NGC 7619 group} 65 Mpc}               &                       & & & HCG 51a                 &   E   &  10.77       &   41.357 $\pm$ .085   \\
NGC 7626                &   E   &  10.94        &   $<$40.68            & & & HCG 51b                 &   S   &  10.34       &   $<$40.973           \\
NGC 7619                &   E   &  10.97        &   $<$40.91            & & & HCG 51c                 &   S0  &  10.73       &   41.017 $\pm$ 0.187  \\
NGC 7617                &   S0  &  9.90         &   40.63  $\pm$ 0.11   & & & HCG 51d                 &   S   &  10.30       &   $<$41.033           \\
PGC 071159              &   S   &  9.72         &   $<$40.14            & & & HCG 51e                 &   E   &  10.53       &   41.017 $\pm$ 0.155  \\
NGC 7623                &   S0  &  10.26        &   $<$39.88            & & & \multicolumn{3}{l}{{\bf~HCG 68} 48 Mpc}               &                       \\
PGC 071120              &   S   &  9.65         &   $<$40.15            & & & HCG 68a                 &   S0  &  10.76       &   41.017 $\pm$ .035   \\
PGC 071085              &   E   &  9.96         &   $<$40.17            & & & HCG 68b                 &   E   &  10.61       &   $<$40.079           \\
NGC 7631                &   S   &  10.24        &   39.75 $\pm$ 0.33    & & & HCG 68c                 &   S   &  10.69       &   40.322 $\pm$ .092   \\
NGC 7611                &   S0  &  10.40        &   40.62 $\pm$ 0.14    & & & HCG 68d                 &   E   &  9.97        &   $<$40.079           \\
PGC 071110              &   S   &  9.61         &   39.93 $\pm$ 0.25    & & & HCG 68e                 &   S0  &  9.72        &   $<$40.079           \\
\multicolumn{3}{l}{{\bf~NGC 7777 group} 133 Mpc}               &                       & & & \multicolumn{3}{l}{{\bf~HCG 90} 53 Mpc}               &                       \\
PGC 072792              &   E   &  10.73        &   $<$41.35            & & & HCG 90a                 &   S   &  10.49       &   40.518 $\pm$ .068   \\
\multicolumn{3}{l}{{\bf~HCG 15} 137 Mpc}               &                       & & & HCG 90b                 &   E   &  10.70       &   40.255 $\pm$ 0.119  \\
HCG 15a                 &   S0  &  10.56        &   40.75 $\pm$ 0.20    & & & HCG 90c                 &   E   &  10.40       &   40.113 $\pm$ 0.171  \\
HCG 15b                 &   S0  &  10.14        &   $<$40.81            & & & HCG 90d                 &   S   &  9.94        &   40.079 $\pm$ 0.156  \\
HCG 15c                 &   S0  &  10.62        &   $<$40.81            & & & \multicolumn{3}{l}{{\bf~HCG 91} 143 Mpc}               &                       \\
HCG 15d                 &   S0  &  10.34        &   41.71 $\pm$ 0.04    & & & HCG 91a                 &   S   &  11.27       &   43.164 $\pm$ .006   \\
HCG 15e                 &   S0  &  10.14        &   $<$40.81            & & & HCG 91b                 &   S   &  10.23       &   40.707 $\pm$ 0.217  \\
HCG 15f                 &   S   &  9.76         &   $<$40.81            & & & HCG 91c                 &   S   &  10.61       &   $<$40.812           \\
\multicolumn{3}{l}{{\bf~HCG 16} 79 Mpc}               &                       & & & \multicolumn{3}{l}{{\bf~HCG 92} 129 Mpc}               &                       \\
HCG 16a                 &   S   &  10.72        &   40.94 $\pm$ 0.07    & & & HCG 92c                 &   S   &  10.76       &   41.262 $\pm$ 0.115  \\
HCG 16b                 &   S   &  10.52        &   40.23 $\pm$ 0.21    & & & HCG 92e                 &   E   &  10.58       &   40.755 $\pm$ 0.249  \\
HCG 16c                 &   S0  &  10.29        &   41.30 $\pm$ 0.04    & & & \multicolumn{3}{l}{{\bf~HCG 97} 131 Mpc}               &                       \\
HCG 16d                 &   I   &  10.30        &   40.72 $\pm$ 0.10    & & & HCG 97a                 &   S0  &  10.69        &   41.97 $\pm$ 0.04    \\
\multicolumn{3}{l}{{\bf~HCG 31} 82 Mpc}               &                       & & & HCG 97b                 &   S   &  10.13       &   $<$40.812           \\
HCG 31a                 &   S   &  9.76         &   $<$40.46            & & & HCG 97c                 &   S   &  10.40       &   $<$40.875           \\
HCG 31b                 &   S   &  9.93         &   $<$40.46            & & & HCG 97d                 &   E   &  10.53       &   40.944 $\pm$ 0.144  \\
HCG 31c                 &   S   &  10.69        &   40.81 $\pm$ 0.08    & & & HCG 97e                 &   S0  &  9.72        &   $<$40.857           \\
HCG 31d                 &   S   &  8.60         &   $<$40.46            & & &                         &       &               &                       \\
\multicolumn{3}{l}{{\bf~HCG 44} 28 Mpc}               &                       & & &                         &       &               &                       \\
HCG 44a                 &   S   &  10.23        &   39.95 $\pm$ 0.08    & & &                         &       &               &                       \\
HCG 44b                 &   E   &  10.35        &   39.65 $\pm$ 0.13    & & &                         &       &               &                       \\
HCG 44c                 &   S   &  9.88         &   $<$39.53            & & &                         &       &               &                       \\
HCG 44d                 &   S   &  9.51         &   $<$39.53            & & &                         &       &               &                       \\
\hline
\end{tabular}
\end{table*}


\section{X-ray properties of Late-type galaxies in Groups}

\subsection{L$_X$:L$_B$\label{splxlb}}

\begin{figure*}
\psfig{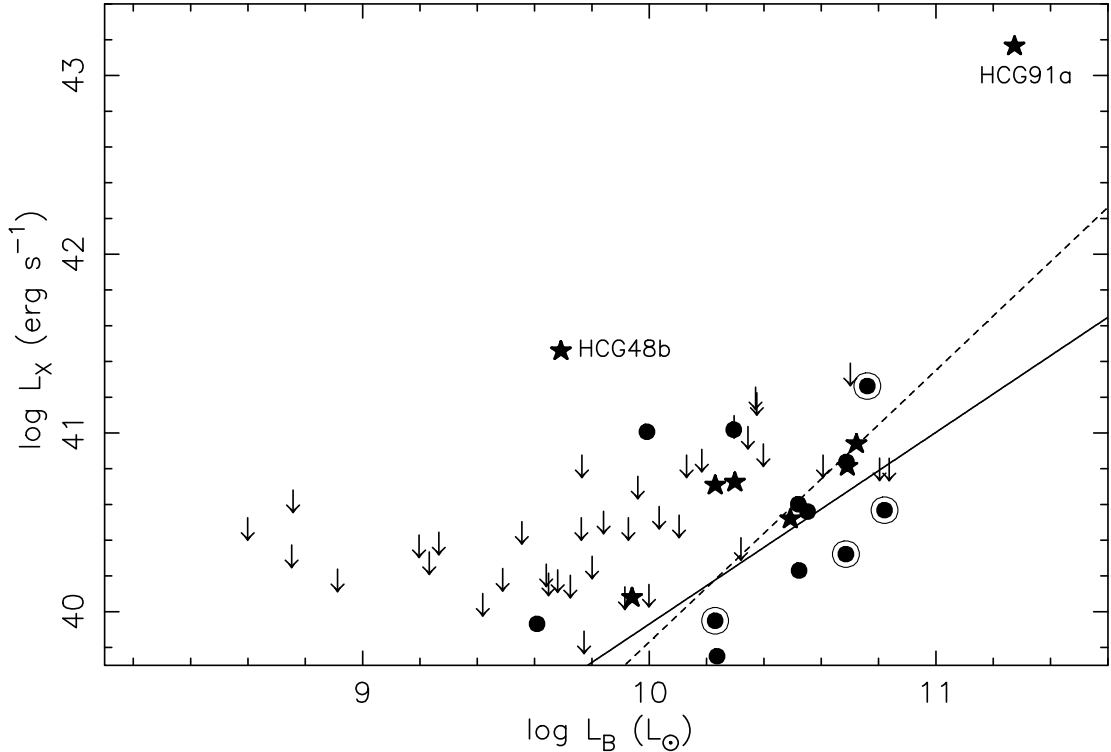}
\caption{\label{fig:spxb}L$_X$:L$_B$ relation for the late-type group
  galaxies. Arrows represent upper limits, other points are detections with
  stars representing likely starbursts (or AGN), ringed circles
  non-starbursts and solid circles late-types whose activity is unknown.
  The solid line is the best fit to all the data, and the dashed line shows
  the relation as derived by \protect\scite{shapley00} for a large sample
  of late-type galaxies observed by {\it Einstein}.}
\end{figure*}    

Figure~\ref{fig:spxb} shows the L$_X$:L$_B$ relation for the late-type
galaxies in our sample. No significant difference in the relation is
apparent between the subsamples in compact and loose groups, so we have
combined the two. It is clear that two points stand out from the trend
described by the rest. The upper of these is HCG 91a, a Seyfert 1 with
powerful central point-like X-ray source associated with the central AGN.
Since this study aims to explore the X-ray properties of {\it normal}
galaxies, we exclude HCG 91a from fits derived below. The second high point
is HCG 48b. This is not known to have an active nucleus, however it is
likely that it contains some sort of active nucleus as it stands out in a
similar way to HCG 91a in both Figures 1 and 2. As a result we also exclude
HCG 48b from the fits derived below (although we will show its effect on
the fits).

As can be seen in Figure~\ref{fig:spxb} there are a large number of upper
limits, so we have used survival analysis to derive best fits to the data.
Survival analysis takes into account both detections and upper limits,
providing that the censoring is random -- i.e. upper limits must be
unrelated to the true values of the parameter being studied. (For example,
this would not be the case if longer observations have been made for
fainter objects). For our sample the luminosity limits are determined by
both exposure time and source distance, and in almost all cases the
galaxies were not the main target of the observation. Note that although
galaxies in more distant groups may have been observed for longer because
of lower fluxes, this does not mean that lower {\it luminosity} galaxies
have been observed longer. Random censoring should therefore be a
reasonable approximation.

Linear regression analysis was carried out using the expectation and
maximisation (EM) and Buckley-James (BJ) algorithms in IRAF. The EM method
requires that the residuals about the fitted line follow a Gaussian
distribution, whilst the BJ algorithm calculates a regression using
Kaplan-Meier residuals and only requires that the censoring distribution
about the fitted line is random. When fitting the lines the uncensored
points were used as independent variables, and the censored points as
dependent variables. In all cases in this paper the two different methods
give similar results, and all quoted lines are based on the mean of these
two methods. A third possible method of linear regression, Schmitt binning,
was not used as the results can be unreliable in cases with heavy
censoring.

The best fit line (solid) shown in Figure~\ref{fig:spxb} is 

\noindent
\begin{math}
\log L_X= (1.07 \pm 0.3) \log L_B + (29.2 \pm 2.1)
\end{math}\\
\noindent
(the slope drops to 0.95 with HCG 48b included). For comparison we also
plot the regression line of \scite{shapley00} who derived the L$_X$:L$_B$
relation for a large sample of spirals (covering a range of environments).

The \scite{shapley00} relationship (slope=1.52$\pm$0.1, intercept=24.6),
whilst steeper, lies within 2$\sigma$ of the slope derived here. It is
therefore not clear that our results are in conflict with those of
\scite{shapley00}, especially since these authors find evidence for extra
emission (possibly from a hot halo) in the largest spirals, which are
mostly absent from our sample.

In galaxy groups it might be expected that galaxy interactions may result
in starbursts which could increase the X-ray to optical luminosity ratio
\cite{read98}. We identify starbursts on the basis of their FIR colours
($f_{60}/f_{100}>0.4$, though note that an AGN may also have such warm FIR
colours -- \pcite{bothun89}). The 12 galaxies for which FIR colour is
available show that the starbursts (or AGN) lie above the best fit line (8
galaxies including HCG 91a and HCG 48b) and the non-starbursts generally
lie below the best fit line (3 out of 4 galaxies).  However, this
difference is not large, and the data generally suggest that some late-type
galaxies in groups may have a small enhancement of their X-ray emission
(relative to the optical) but overall these galaxies follow the same
L$_X$:L$_B$ relation as galaxies in other environments.

\subsection{L$_X$:L$_{FIR}$}

\begin{figure*}
\psfig{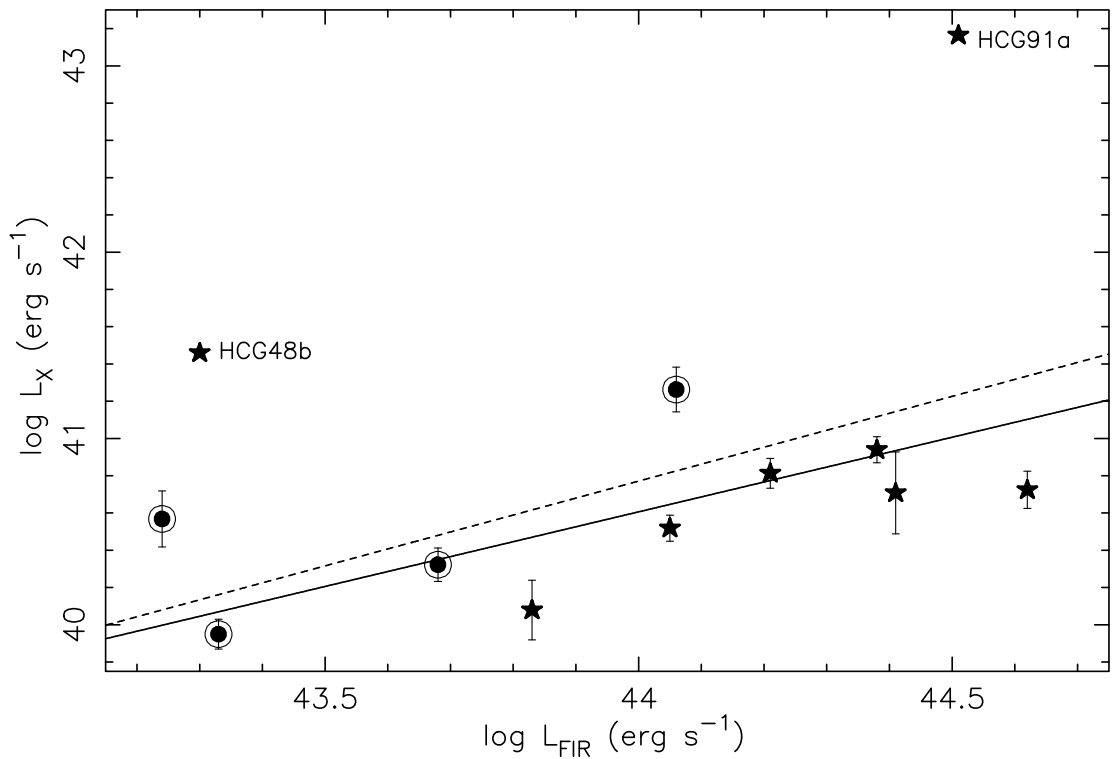}
\caption{\label{fig:spxfir}Late-type galaxy $L_X$:$L_{FIR}$ relation. Only
  galaxies detected in the X-ray and with determined far infrared
  luminosities are plotted. Stars represent starburst galaxies or AGN, and
  ringed circles non-starbursts. The overlaid lines are taken from fits to
  the spiral samples of \protect\scite{read00} (solid line) and
  \protect\scite{shapley00} (dashed).}
\end{figure*}

In Figure~\ref{fig:spxfir} we plot the L$_X$:L$_{FIR}$ relation for the
late-type galaxies in our sample. FIR luminosities are calculated from
IRAS 60 and 100 $\mu m$ fluxes using \\

\noindent
\begin{math}
L_{FIR} = 3.65 \times 10^5 (2.58 S_{60\mu m} + S_{100\mu m}) D^2 ~~~~(L_{\odot})
\end{math}\\

\noindent
(e.g. \pcite{devereux89}) where $D$ is the distance in Mpc, $S_{60\mu m}$
and $S_{100\mu m}$ are the IRAS 60 and 100 $\mu m$ fluxes in Janskys. Most
of the galaxies with IRAS fluxes recorded are in the Hickson Compact
groups, and the flux values are taken from \scite{hickson89}. For other
galaxies, IRAS fluxes are taken from NED. Only galaxies detected in the
X-ray and having a determined FIR luminosity are plotted. The dashed line
in Figure~\ref{fig:spxfir} is from \scite{shapley00} and the solid line is
from a smaller ROSAT PSPC survey of nearby spirals by \scite{read00}.

The rather tight correlation between the two variables, which has been
noted previously, is clearly apparent. The two points well away from the
trend are once again HCG 91a and HCG 48b. This strongly suggests that HCG
48b, like HCG 91a, contains an AGN, since strong interaction-induced
starburst activity tends to {\it decrease} the ratio of L$_X$/L$_{FIR}$
\cite{read98}. In fact the \scite{read00} line in Figure~\ref{fig:spxfir}
is a good fit to both normal and starburst galaxies, with starbusts
occupying the high L$_X$ (or high L$_{FIR}$) end of the relation. The
starbursts as identified in \S~\ref{splxlb} above do indeed tend to lie in
this region of the relation.  Despite the limited number of data points it
can be seen that most late-type galaxies in groups again follow the same
relation as spirals in other environments.


\section{X-ray properties of Early-type galaxies in groups}

\begin{figure*}
\psfig{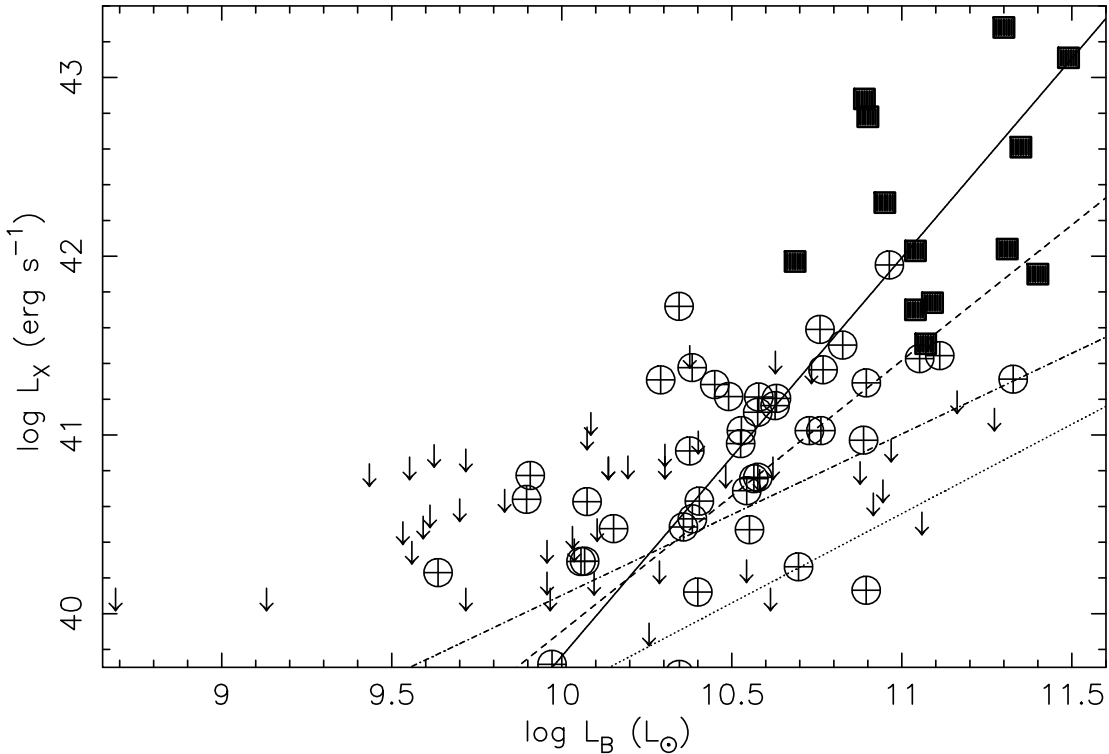}
\caption{\label{fig:earlyxb}L$_X$:L$_B$ relation for the early-type group
  galaxies. The filled squares represent the central-dominant group
  ellipticals, the circles with crosses represent other detected early-type
  galaxies in groups, and the arrows are upper limits. Also plotted on the
  graph are the following lines, the L$_X$:L$_B$ relation as derived by
  Beuing et al. (1999 - solid line), an estimate of the discrete source
  contribution (dotted line), the best fit to the full group galaxy sample
  (dashed line) and the best fit to the group sample excluding central
  dominant galaxies (dot-dash line).}
\end{figure*}
\nocite{beuing99}

Figure~\ref{fig:earlyxb} shows the L$_X$:L$_B$ relation for the early-type
galaxies in our sample. No significant difference is apparent between the
compact and loose group samples. The crossed circles represent the
non-central group galaxy detections and the arrows upper limits. Shaded
squares are central-dominant group galaxies, as defined in \S~\ref{cenred}
and include HCG 97a, which is a central-dominant galaxy from the HRI sample.
Also plotted for comparison are the best fit to the early-type galaxy
L$_X$:L$_B$ relation as determined by \scite{beuing99}, on the basis of
their study of a large sample of early-type galaxies derived from the ROSAT
All Sky Survey (RASS), and an estimate of the discrete source contribution.
This latter estimate was derived using the mean hard component L$_X$/L$_B$
derived by \scite{matsushita00}. These authors used spectral fits to ASCA
data to calculate the contribution of the hard spectral component,
identified with discrete galactic sources, in the 0.5-10~keV band, assuming
a 10~keV thermal bremsstrahlung model for its spectrum. As we use a 1~keV
\scite{raymond77} model for all our galaxies we have converted the
\scite{matsushita00} L$_X$/L$_B$ ratio to our assumed model and bandpass to
give a discrete source estimate of L$_X$/L$_B$=29.56. This value is in
reasonable agreement with a number of other estimates of this hard discrete
source contribution, once the effects of differing bandpasses and models
are taken into account \cite{canizares87,ciotti91,matsumoto97,irwin98}.  It
can also be seen to constitute a reasonable lower envelope for the X-ray
luminosity of the galaxies in our sample. It has been suggested that there
is also a very soft component associated with discrete sources (e.g.
\pcite{irwin98}). If this component is present in all galaxies its effect
on our estimated discrete source contribution would be somewhat dependent
on the absorbing column, but could typically increase it by a factor of
$\sim$2. However, recent Chandra observations have resolved much of the
discrete source contribution in early-types \cite{sarazin00,blanton00}, and
Chandra spectral analysis of this discrete component show that any soft
component is very weak relative to the hard component \cite{blanton00}.

As described in \S~\ref{splxlb}, we again use survival analysis to derive
fits to the data, including the upper limits available. Fitting to the
whole dataset gives\\

\noindent
\begin{math}
  \log L_X = ( 1.5 \pm 0.2 ) \log L_B + ( 24.7 \pm 2.0 )
\end{math}\\

\noindent
which is plotted as the dashed line in Figure~\ref{fig:earlyxb}. Although
this fit is flatter than the \scite{beuing99} line (slope=2.23 $\pm$ 0.12),
a large fraction of our data lie in the region log~$L_X \leq$~40.5. Below
this luminosity previous work indicates that the slope of the relation is
approximately unity \cite{eskridge95}, we therefore expect a somewhat
flatter slope than derived by \scite{beuing99}, who had many more luminous
galaxies in their sample. In fact if we restrict the fit to the more
optically luminous galaxies the slope does indeed steepen to $\sim$ 2.2
(and if we restrict the \scite{beuing99} data to the less optically
luminous galaxies their slope flattens).

However it is clear that the central-dominant group galaxies all lie in the
upper right region of the graph. This along with their central position in
the group suggests that they may not be typical of other early-type group
galaxies. Thus we also fit a regression line to the data after excluding
all central-dominant galaxies, obtaining a best fit of,\\

\noindent
\begin{math}
  \log L_X = ( 0.90 \pm 0.18 ) \log L_B + ( 31.1 \pm 1.8 )
\end{math}\\

\noindent
which is significantly flatter than the previous fit and is plotted as the
dot-dash line in Figure~\ref{fig:earlyxb}.

While the fit to the full dataset does appear to be consistent with
previous work, of more interest is the very flat slope obtained if the
central-dominant group galaxies are excluded. Since this line has a slope
consistent with unity, one interpretation might be that the X-ray emission
is primarily from stellar sources, and that these non-central group
galaxies do not contain a significant hot halo. However the level of this
line is a factor $\sim2.5$ above the expected hard discrete source
contribution, and even if a soft component is included in the discrete
source contribution, many of these galaxies will still lie above the line,
suggesting that a hot halo still remains. We will return to this issue
below.

A further factor which might be influencing our result, which is
significantly flatter than that derived in previous studies (e.g.
\pcite{beuing99,eskridge95}), is that these other studies included galaxies
from a range of environments, whilst we have specifically targeted
galaxies in X-ray bright groups. To explore this, we now compare group
galaxies to those in low density environments.

\section{Comparison with non-group early-type galaxies}

\begin{table*}
\hspace{-1.5cm}
\begin{center}
\begin{tabular}{lccc}

Dataset & Slope & Intercept & no. galaxies\\
\hline
Field galaxies & 2.1 $\pm$ 0.4 & 17.8 $\pm$ 4 & 64(11)\\
Group galaxies (only BGGs) & 3.5 $\pm$ 0.6 & 2.6 $\pm$ 7 & 50(24) \\
Group galaxies (no BGGs) & 2.1 $\pm$ 0.2 & 18.2 $\pm$ 4 & 84(17) \\
Group galaxies with emission $\geq$ 200kpc & 2.7 $\pm$ 0.3 & 11.4 $\pm$ 4 & 25(25) \\
Group galaxies with emission $<$ 200kpc & 1.4 $\pm$ 0.4 & 25.0 $\pm$ 4 & 109(16) \\
\hline
\end{tabular}
\end{center}
\caption{\label{tab:beuingfits}Results of survival analysis fitting on each 
  of the environmental subsets of the \protect\scite{beuing99} sample. The
  first number in the 4th column is the total number of galaxies, the
  number in the brackets is the number of detections.}
\end{table*}

To investigate the effects of environment on the properties of early-type
galaxies we adopt two approaches. Firstly we use a previous large study of
the X-ray properties of early-type galaxies.  This large sample will be
used to derive group and field subsamples, and the properties of each
subsample examined. The second approach involves defining a sample of very
isolated early-type galaxies whose properties we can compare with our group
galaxies.  This latter approach results in a smaller sample, but with much
more reliable X-ray parameters. Both approaches are described in detail
below.


\subsection{Environmental effects in the RASS sample\label{prev_work}}

The largest ROSAT study to date of early-type galaxies is that of
\scite{beuing99}, with nearly 300 galaxies examined using X-ray data from
the RASS. However one complication in the Beuing et al. L$_X$ values is the
presence of intragroup or intracluster X-ray emission which Beuing et al.
associate with the galaxy if it lies at the centre of the extended X-ray
emission.  This will tend to increase the quoted L$_X$ for high luminosity
galaxies (which are especially likely to lie in clusters, or in the cores
of galaxy groups), increasing the slope of the L$_X$/L$_B$ relation at high
L$_B$ values.

In order to investigate any possible environmental dependence in the Beuing
\etal sample we divided the sample into a number of group and field
subsamples. Cluster galaxies were identified and removed based on
membership in the \scite{abell89} and \scite{faber89} catalogues. The X-ray
properties of cluster galaxies determined by Beuing et al. may be severely
contaminated by cluster emission, and will be best addressed by future
observations with the high spatial resolution of Chandra. These are not the
object of the present study. Group membership is taken from the all sky
group catalogue of \scite{garcia93}, which has a velocity limit of 5,500 km
s$^{-1}$. Any galaxy not classified as a member of a cluster or group from
these catalogues is considered to be in the field. Galaxies more distant
than 5,500 km s$^{-1}$ are removed to avoid possible misclassification.
This process produces 127 group galaxies and 81 in the field. To explore
any possible difference between central-dominant group galaxies and other
galaxies in groups, we further split the Beuing et al. group sample into a
sample of BGGs and one with all BGGs removed.

\scite{beuing99} record the radii to which they detect emission around each
galaxy. In an attempt to explore the effects of contamination of their
galaxy fluxes by surrounding group emission, we also split the group sample
into subsamples with extents greater than and less than 200~kpc. Galaxies
with extents $>200$~kpc are very likely to include a significant amount of
group emission.

We fitted the L$_X$:L$_B$ relation for each subsample separately using
survival analysis techniques as described earlier. The results for the
slopes of these fits for non-cluster galaxies are shown in
Table~\ref{tab:beuingfits}. These results do appear to indicate the
presence of significant environmental effects, with the BGG galaxies having
a very steep slope, the field sample a flatter relation, and group galaxies
with emission confined within 200~kpc a flatter slope still. However some
of these fits may be biased, since some of the subsets have very few
detections and many upper limits.

\begin{figure*}
  \psfig{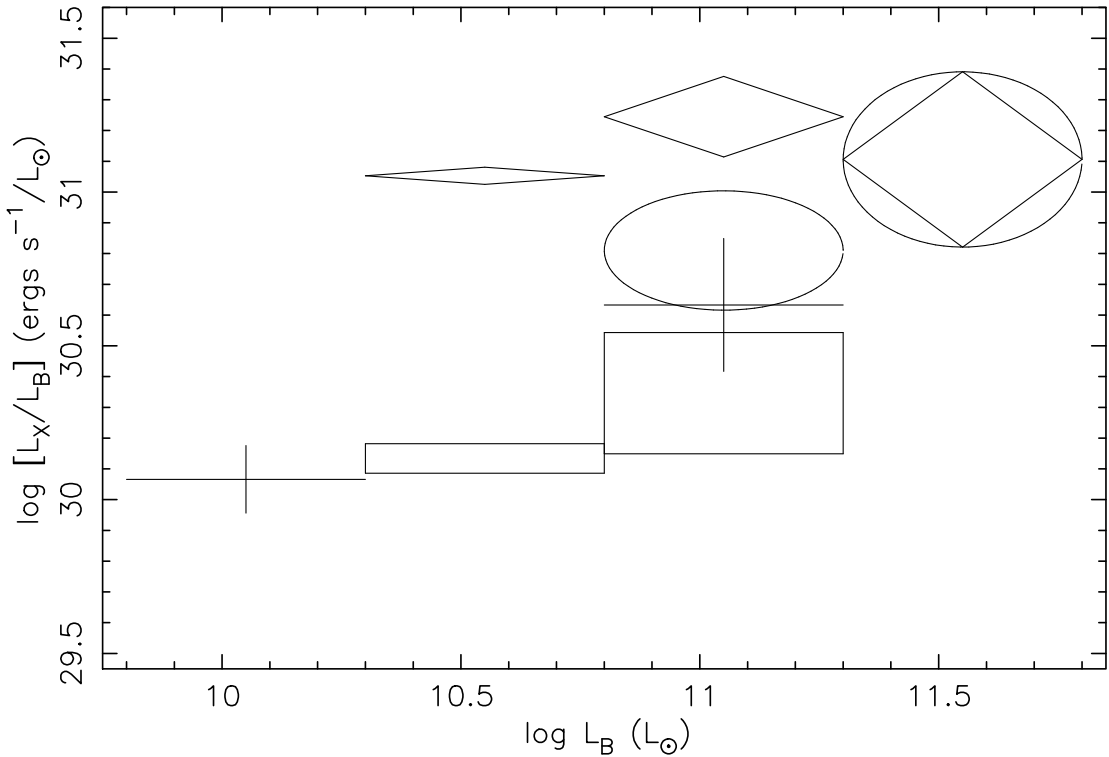}
\caption{\label{fig:ratio1}L$_X$/L$_B$ versus L$_B$ for the binned data
  from the \protect\scite{beuing99} subsets. Ellipses represent BGGs,
  crosses non-BGG group galaxies, and light boxes field galaxies. Diamonds
  are group galaxies with emission recorded as being $>$200 kpc in extent.}
\end{figure*}

In order to reduce any bias in our comparisons we binned our data into
optical luminosity bins and used the Kaplan-Meier estimator (which includes
upper limits) within \textsc{iraf} to derive the mean value of L$_X$/L$_B$
for each bin.  This technique works by redistributing the contribution from
each upper limit over all the lower points within a bin.  One effect of
this, is that estimates may be seriously biased where all the lowest values
within a bin consist of upper limits. To avoid this problem, we discard
bins which have many more upper limits than detections. In
Figure~\ref{fig:ratio1} we plot log(L$_X$/L$_B$) versus log L$_B$ for the
subsets of the \scite{beuing99} data. There is some evidence from this plot
that the BGGs (which often have r$_{ext}>200$kpc in \pcite{beuing99}) are
more X-ray luminous in general than the other early-type galaxies in
groups, or the field galaxies, though the comparison is hampered by a lack
of detections for non-BGG group galaxies. The rather high non-BGG point at
log(L$_B)=$11.05 includes only 4 detections all of which show emission
extending to a radius well over 200~kpc. This suggests that the fluxes for
these galaxies may include a significant amount of group emission.

The field galaxy sample shown in Figure~\ref{fig:ratio1} could also be
contaminated. The `field' designation here relies on galaxies not appearing
in the group catalogue of \scite{garcia93}. It could therefore include
galaxies which are not all that isolated. In addition, the field sample
could be contaminated with `fossil groups'
\cite{ponman94,mulchaey99a,vikhlinin99}. These would tend to lie in the
high L$_B$ region of the plot, as they are thought to be the result of a
number of galaxy mergers within a compact group, thus leaving a single
bright early-type in the centre of the group potential, with no other
bright galaxies nearby. Such systems can have very high X-ray luminosities,
arising from the hot intergalactic medium of the erstwhile group.

The conclusion from this section is that there does appear to be some
evidence for environmental effects within the \scite{beuing99} sample, but
that the statistics are limited, the quality of the L$_X$ values
questionable, and there is some doubt about some of the environmental
classification of some galaxies. To address these concerns, we now attempt
to identify a sample of genuinely isolated galaxies with available X-ray
data, which can be compared with our group sample.

\subsection{Isolated galaxies compared with group galaxies}
\label{sec:red:iso}

First we need to define a sample of isolated early-type galaxies. This was
done using the Lyon-Meudon Extragalactic Data Archive (LEDA). This
catalogue contains information for about 100,000 galaxies, of which
$\sim$40,000 have enough information recorded to be of use to us. From this
sample we select galaxies which fit the following criteria:
\begin{itemize}
\item Morphological type $T\leq-3$.
\item Virgo corrected recession velocity $V\leq9000$ km s$^{-1}$.
\item Apparent magnitude $B_T\leq14.0$.
\item Not listed as a member of a Lyon Galaxy Group \cite{garcia93}.
\end{itemize}

The restrictions on apparent magnitude and recession velocity are imposed
to minimize the effect of incompleteness in the catalogue. The LEDA
catalogue is known to be 90\% complete to $B_T=14.5$ \cite{amendola99}, so
our sample should be close to being statistically complete.

This selection gives 330 galaxies which can be considered as potential
candidates. These are compared to the rest of the catalogue and accepted as
being isolated if they have no neighbours which are:
\begin{itemize}
\item within 700 km s$^{-1}$ in recession velocity.
\item within 1 Mpc in the plane of the sky.
\item less than 2 magnitudes fainter in $B_T$.
\end{itemize}

These criteria are imposed to ensure that galaxies do not lie in groups or
clusters and that any neighbours they do have are too small to have had any
significant effect on their evolution and X-ray properties.

To check the results of this process, all galaxies are compared against NED
and the Digitized Sky Survey (DSS).  A NED search in the area within 1~Mpc
of the galaxy picks out galaxies which are not listed in LEDA, and
examination of the DSS reveals any galaxies of similar brightness to the
candidate which are not listed in either catalogue. We believe that these
measures ensure that the isolation of the galaxy is real, and not produced
by errors or omissions in the catalogue.  Our final sample consists of 36
isolated early-type galaxies.

\begin{table*}
\hspace{-1.5cm}
\footnotesize
\begin{center}
\begin{tabular}{lcccccccc}

Name & Distance & Hubble & $B_T$ & D$_{nn}$ & X-ray Data & log L$_X$ &
$L_X/L_B$ & notes\\
 & (km s$^{-1}$) & Type & (mag) & (Mpc) &    & (erg s$^{-1}$) & & \\
\hline
\hline
NGC 821 & 1747 & -4.8 & 11.74 & $>$1 & RP & $<$40.94 & $<$30.25 & RASS\\
NGC 1132 & 6904 & -4.9 & 13.26 & $>$1 & A & $\sim$43.0 & $\sim$31.84 & MZ99\\
NGC 2110 & 2091 & -3.0 & 13.40 & $>$1 & A & 42.1 & 32.04 & Seyfert 2\\
NGC 2271 & 2412 & -3.2 & 13.24 & $>$1 & RP & $<$41.12 & $<$30.43 & RASS\\
NGC 2865 & 2451 & -4.1 & 12.41 & $>$1 & RP & $<$40.76 & $<$30.22 & RASS\\
NGC 4555 & 6775 & -4.8 & 13.31 & $>$1 & RP & 41.15 & 30.03 & Satellites\\
NGC 6776 & 5281 & -4.1 & 13.01 & $>$1 & RP & 41.18 & 30.15 & post-merger\\
NGC 7796 & 3145 & -3.9 & 12.42 & $>$1 & RP & 41.58 & 30.76 & \\
\hline
 & & & & & & & & \\
\hline
NGC 993 & 6967 & -3.4 & 14.64 & 0.67 & RP & $<$41.77 & $<$30.14 & \\
NGC 2418 & 5066 & -4.9 & 13.34 & 0.4 & RP & 41.36 & 30.49 & \\
IC 3171 & 7096 & -3.3 & 14.60 & 0.4 & RP & $<$41.40 & $<$30.75 & \\
CGCG 196-012 & 8897 & -3.0 & 15.52 & 0.53 & RP & $<$41.92 & $<$31.34 & \\
MCG +3-47-10 & 5249 & -3.0 & 14.78 & $>$1 & E & $<$41.97 & $<$31.65 & \\
MCG +5-31-79 & 7522 & -3.7 & 15.31 & 0.55 & A & $<$42.73 & $<$32.31 & \\
MCG +5-31-151 & 7153 & -3.8 & 14.42 & $>$1 & A & $<$41.95 & $<$31.21 & \\
\hline
\normalsize
\end{tabular}
\end{center}
\caption{\label{tab:iso}Isolated Early-type Galaxies with X-ray data. E is 
  {\it Einstein}, A is {\it ASCA}, RP is {\it ROSAT}
  PSPC. $D_{nn}$ is the distance in the plane of the sky to the
  nearest neighbour of similar magnitude and recession
  velocity. The $L_X$ value for NGC 1132 is from Mulchaey \& Zabludoff (1999)
  and values for NGC 821, NGC 2271 and NGC 2865 are taken from the {\it
  ROSAT All-Sky Survey} sample of Beuing et al. (1999). $L_X$ for NGC
  2110 is taken from Boller et al. (1992). All $L_X$ values apart from
  those of NGC 6776, NGC 1132 and NGC 2110 are based on fits with fixed
  0.25 solar abundance and 1keV temperature.}
\end{table*}

Comparison of these 36 galaxies with previously published work, and with
pointed X-ray observations ({\it ROSAT}, {\it ASCA} \& {\it Einstein})
reveals that only 8 have X-ray data available.  Of these, two (NGC 821 and
NGC 2271) have been observed in the ROSAT All-Sky Survey (RASS), but were
not detected at 3$\sigma$ confidence, and have no pointed observations
available. NGC 6776 is a fairly recent post-merger \cite{sansom88} as is
NGC 2865 \cite{hau99}, NGC 2110 is a Seyfert galaxy, and NGC 1132 is a
possible ``fossil'' group elliptical \cite{mulchaey99a}.  There is also a
possible problem with NGC 4555, in that the DSS shows several small
galaxies projected near it.  Although they are probably not large enough to
exclude NGC 4555 from consideration, their possible effect on the galaxy
should not be ignored.

To expand and improve this subset, we chose to relax the isolation
conditions and the apparent magnitude cut. We thus include galaxies that
are either less isolated (no neighbours within 0.4 Mpc in plane of sky) or
which are faint enough for their isolation to be uncertain ($B_T>14$).
Although these do not fit our rigorous criteria, we believe that they are
isolated enough to provide useful extra information. The relaxed criteria
should still exclude galaxies in truly dense environments, but might
include galaxies in loose associations or in the outer fringes of large
clusters. Once again we searched for X-ray data from either {\it ROSAT},
{\it ASCA} or {\it Einstein}.

Galaxies defined as isolated by the rigorous definition are listed in the
top half of Table~\ref{tab:iso}, and additional galaxies are given in the
lower half. X-ray luminosities for these galaxies were derived using a
simple standard reduction. The instrument from which the data were obtained
for each galaxy is listed in Table~\ref{tab:iso}. Luminosities from {\it
  ROSAT} were background subtracted and corrected. Spectra were extracted
from an aperture of radius $\sim$6 effective radii. Detections and upper
limits were derived using a Raymond-Smith plasma \cite{raymond77} model
with hydrogen column fixed at Galactic values \cite{stark92}, temperature
fixed at 1 keV and 0.25 solar metallicity.

\nocite{tully88}
\nocite{mulchaey99a}
\nocite{beuing99}
\nocite{boller92}

Analysis of {\it ASCA} data was carried out using the screened data sets
available from LEDAS (Leicester Database and Archive Service). Spectra were
extracted using the \textsc{xselect} package in \textsc{ftools} and fitted
using \textsc{xspec}.  In cases where no reasonable fit could be obtained
and no source was obvious, we used the count rate within the circular
aperture to estimate a 3$\sigma_b$ upper limit on the galaxy luminosity.
Conversion of count rate into flux was performed using \textsc{w3pimms}.
Analysis of the {\it Einstein} data was performed using a reduced dataset
available on CDROM from the Smithsonian Astrophysical Observatory. This did
not give a detection, and a 3$\sigma_b$ upper limit was derived based on
the number of counts found in a 6 r$_e$ radius around the galaxy position.

\begin{figure*}
\psfig{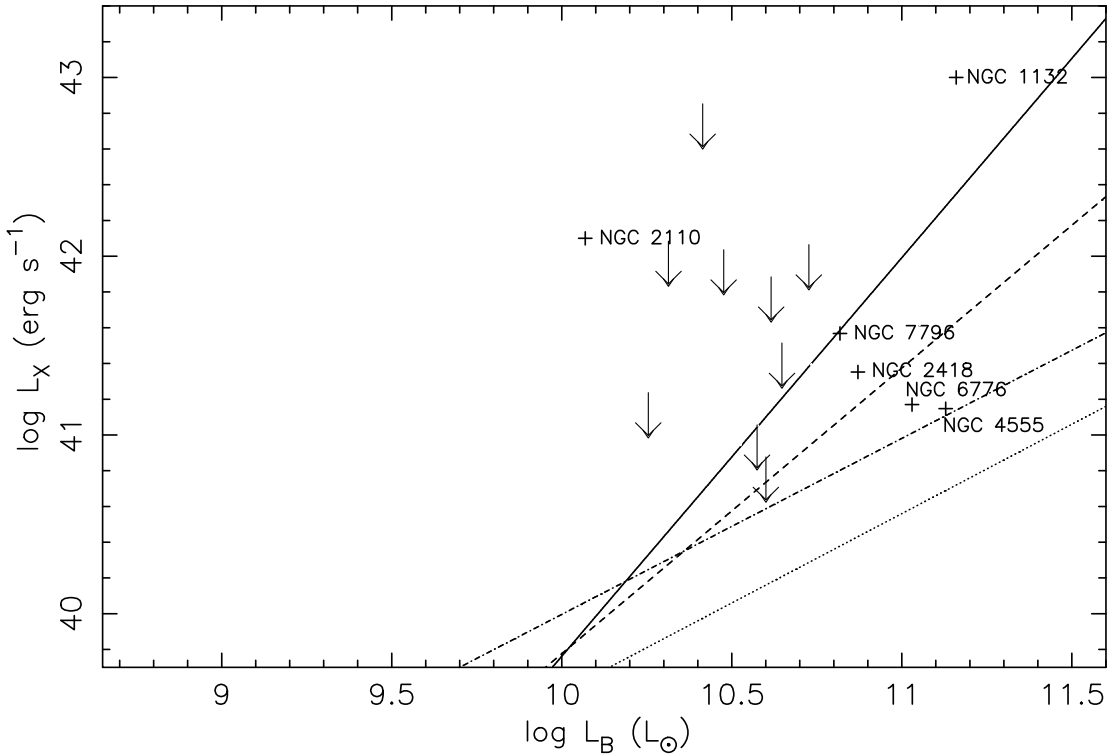}
\caption{\label{fig:isoxb}L$_X$:L$_B$ relation for our sample of isolated
  early-type galaxies. The labeled crosses mark the detections and the
  arrows the upper limits. The lines plotted are the same as those in
  Figure~\ref{fig:earlyxb}.}
\end{figure*}

The L$_X$:L$_B$ values for the isolated galaxies are plotted in
Figure~\ref{fig:isoxb}. Also plotted for comparison are the lines shown in
Figure~\ref{fig:earlyxb}. The crosses mark detections, and the
corresponding galaxy names are given on the plot.  Upper limits are denoted
by arrows.  One point of interest is that the ``fossil'' group elliptical
(NGC 1132) lies in the same region of the plot as the central-dominant
group galaxies.  Note that NGC 2110 is a Seyfert galaxy.

\begin{figure*}
  \psfig{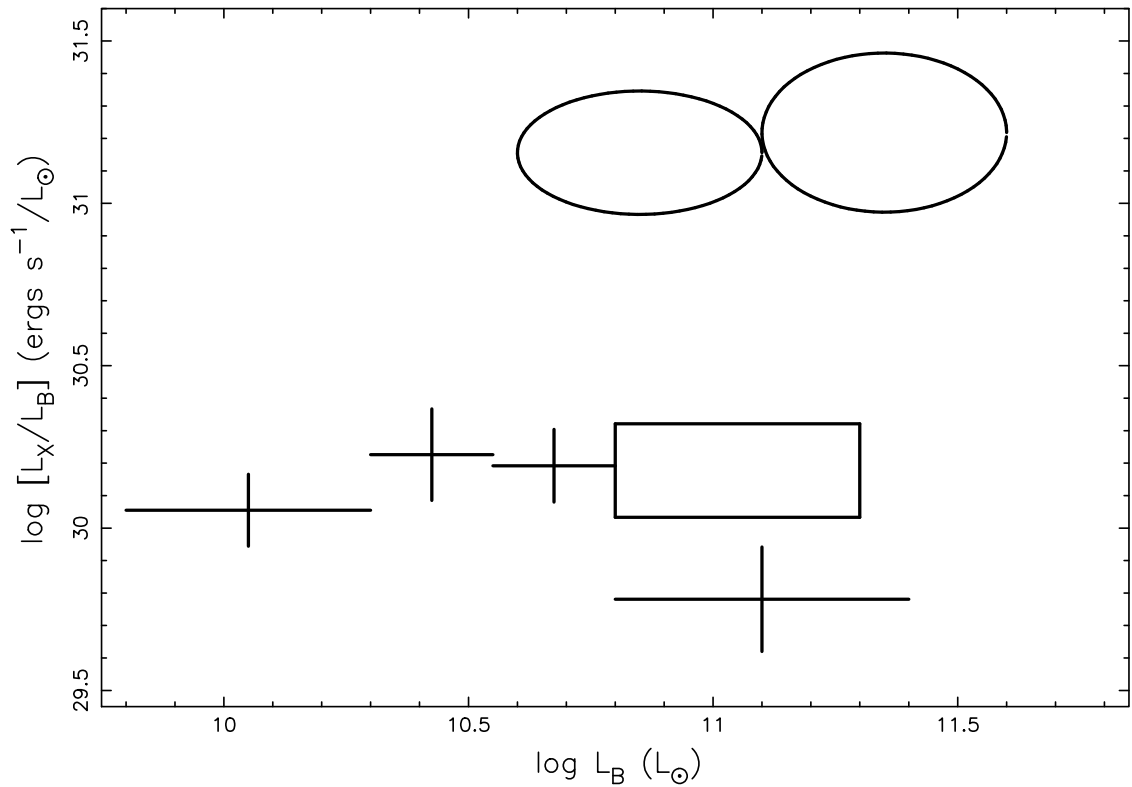}
\caption{\label{fig:ratio2}L$_X$/L$_B$ versus L$_B$ for the binned data 
  for group and isolated early-type galaxies derived in this paper.
  Ellipses represent central-dominant group galaxies, crosses non-central
  group galaxies and the box isolated galaxies.}
\end{figure*}

As can be seen, the small numbers of isolated galaxies do not provide
strong constraints on the L$_X$:L$_B$ relation. As a result we now bin up
the data for the isolated galaxies, and for the group galaxy data derived
in this paper. We have binned these in the same way as for the Beuing
subsets in \S~\ref{prev_work}. Figure~\ref{fig:ratio2} shows
log(L$_X$/L$_B$) versus log L$_B$ for the group dominant galaxies,
non-central galaxies and isolated galaxies. Due to their special nature, we
have excluded the fossil group and the Seyfert from the isolated galaxy
sample. No bins in the group samples are omitted due to poor statistics
this time. This Figure shows a much clearer picture than the results
derived above from the Beuing sample. The L$_X$/L$_B$ ratio for non-central
group galaxies and isolated galaxies is consistent with a constant log
L$_X$/L$_B$ value of $\approx 30$~(erg~s$^{-1}L_{\odot}^{-1}$), whilst
the central dominant group galaxies are clearly far more X-ray luminous
than other galaxies of similar optical luminosity.

The significant X-ray over-luminosity of the central group galaxies
relative to the others, suggests that some process is enhancing the X-ray
emission from these central galaxies, whilst other early-type galaxies in
groups are similar in their X-ray properties to isolated galaxies.


\section{Discussion}
\label{sec:dis}

There are two main results from the data presented above. Firstly, the
X-ray properties of spirals in galaxy groups appear to be indistinguishable
from spirals in other environments. Secondly, the X-ray properties of
early-type galaxies in groups appear to depend on whether the galaxy is
centrally located or not.

The properties of the late-type galaxies in groups are perhaps
unsurprising, given that in late-type galaxies much of the X-ray emission
is believed to come from stellar sources.  Spirals which lie significantly
above the L$_X$:L$_B$ relation most likely contain active nuclei, and also
show excess X-ray emission relative to the L$_X$:L$_{FIR}$ relation.  The
solid line plotted in Figure~\ref{fig:spxfir} is a good fit to both normal
and starbursting galaxies \cite{read00} suggesting that the excess X-ray
emission seen in HCG 91a and HCG 48b is not due to starbursts.

In the case of compact groups, there is strong evidence that the group
environment leads to galaxy interactions \cite{deoliveira94}, but the
statistical evidence that this leads to increased starburst activity is
controversial (see e.g. \pcite{hickson97,shimada00}). Our results do not
throw any additional light on this debate, since we do not have a
well-controlled statistical sample. Rather, the implication of our results
is that the X-ray luminosity of group galaxies, arising from a combination
of discrete sources and starburst-related hot gas (see \pcite{read98}),
scales with galaxy size and starburst activity, in a way which appears
indistinguishable from field spirals.

In the case of early-type galaxies, our results imply that these need to be
separated into two classes -- central-dominant galaxies, and others -- and
that the failure to recognise this distinction has compromised much
previous work on the X-ray properties of early-type galaxies.

What is the origin of this distinction in properties?  Central-dominant
group galaxies appear to be located at the centre of the group potential
both in radial velocity and on the sky \cite{zabludoff98}. This suggests
that these galaxies are not moving significantly with respect to the group
potential, in contrast to other group galaxies. A galaxy moving through the
intragroup medium may undergo ram pressure stripping \cite{gunn72} or
viscous stripping \cite{nulsen82}. Simulations of galactic halos in
clusters also suggest that tidal stripping by the cluster potential can
remove a significant amount of a galaxy halo \cite{okamoto99}. All these
processes could reduce the amount of hot gas that the galaxy could retain,
thus reducing the X-ray luminosity towards the value expected from stellar
sources alone. In contrast, a centrally located, stationary galaxy may be
able to accrete additional gas from the surrounding group potential,
enhancing its X-ray luminosity.

In the absence of active stripping, the hot gas content of an early-type
galaxy is determined by the balance between gas loss from stars, gas
infall from the surroundings, and energy sources (notably type Ia
supernovae) within the galaxy which control the energetics of the gas. The
result can be a hydrostatic halo with a cooling flow, a global outflow, or
a `partial wind' in which an outer wind region surrounds a central inflow
\cite{ciotti91,pellegrini98,brighenti99}.  In the case of global inflows,
the X-ray luminosity is essentially equal to the whole of the injected
supernova luminosity (and even higher if gas flows in from outside the
galaxy), whilst in the case of winds, much of this energy may be lost in
kinetic energy, and L$_X$ is typically much lower.

Previous discussion of the observed L$_X$:L$_B$ relation in terms of such
models, has started from the basis that this relation is steep
(approximately L$_X \propto $L$_B^2$) for optically bright early-type
galaxies.  \scite{brown00}, for example, suggest that the observed
L$_X$:L$_B$ relation may be explained by a model based on the transition
between galaxies with total winds, partial winds and those in which the gas
is retained. They calculate that this transition should occur over the
range 10.7 $< \log L_B <$ 11.2, leading to a rise in L$_X$/L$_B$ across
this range. In a galaxy group the surrounding intragroup medium might be
expected to help suppress galaxy winds, pushing the region of steep
L$_X$/L$_B$ to lower luminosities. However for our sample of
non-central-dominant group galaxies, the average ratio of L$_X$/L$_B$ is
approximately constant over the range 9.8 $< \log L_B <$ 11.3.

What of the effects of environment on early-type galaxies?  Our results
(Figure \ref{fig:ratio2}) show no significant difference between isolated
galaxies, and non-central group galaxies. In contrast, the studies of
\scite{white91} and \scite{brown00} both found that the X-ray properties of
early-type galaxies have some dependence on environment, but in opposite
senses. \scite{brown00} find that on average, galaxies in denser
environments (as defined by the Tully volume density parameter $\rho$) are
more X-ray luminous. In fact they see that galaxies in low density
environments have low L$_X$/L$_B$, whilst those in denser environments show
a larger scatter in L$_X$/L$_B$.  In the light of our results, it is
natural to ask whether this result might be entirely due to the inclusion
of central-dominant group galaxies in the \scite{brown00} sample. In the
\scite{beuing99} sample, only 3 out of 58 galaxies with log L$_B >$11.0 are
found in the field. This suggests that almost all optically bright
early-type galaxies are located in groups and clusters. Many of these
brightest galaxies will be central-dominant group galaxies, which will
naturally lie in rather dense environments, and have a high value of
L$_X$/L$_B$.

\begin{figure*}
\psfig{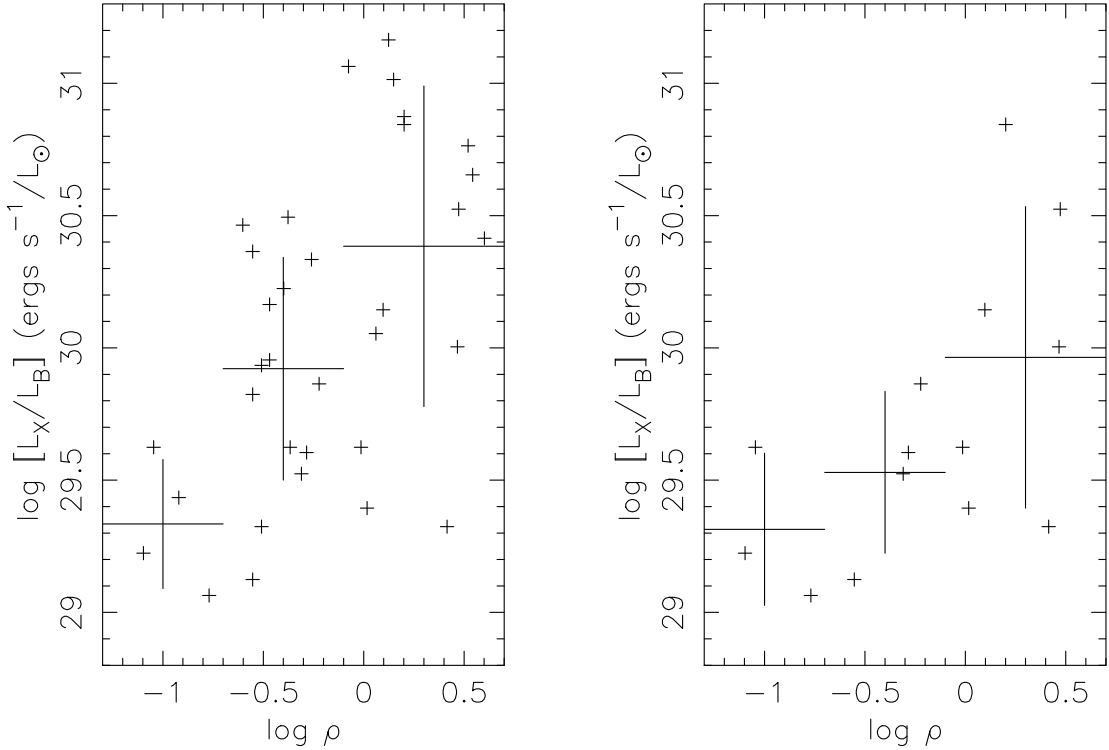}
\caption{\label{fig:bb00}On the left is the original data from
  \protect\scite{brown00} showing a trend of increasing L$_X$/L$_B$ with
  galaxy volume density, $\rho$. On the right is the same relation but with
  all brightest group galaxies removed. On both plots the small crosses are
  galaxy data points and large crosses are binned up data.}
\end{figure*}

If we remove all the brightest group galaxies from the \scite{brown00}
sample (optically brightest as defined by \pcite{garcia93} plus two from
\pcite{helsdon00}) the trend in L$_X$/L$_B$ is considerably reduced. If
galaxies identified as being at the centre of sub-clumps within Virgo
\cite{schindler99} are also removed (they may represent the cores of groups
which have fallen in) the remaining data show only a weak trend ($\sim
1.5\sigma$) with galaxies in the densest environments showing a lot of
scatter, while galaxies in other environments show an almost constant
L$_X$/L$_B$. This is shown in Figure~\ref{fig:bb00}, which contrasts the
original \scite{brown00} dataset with the same data with central group
galaxies removed (note that an approximate conversion to our bolometric
X-ray bandpass have been carried out on the data). The two highest points
remaining within the high density bin are in clusters (Virgo and Fornax),
and with the exception of these, the data show that non-central-dominant
group galaxies across a range of environments have L$_X$/L$_B$ ratios with
rather little variation (scatter is factor of $\sim$ 3 about a mean value
of log L$_X$/L$_B$ $\approx 29.5$ ergs s$^{-1}$ L$_{\odot}^{-1}$ - we
believe this value is lower than our value of log L$_X$/L$_B$ $\approx
30$~(erg~s$^{-1}L_{\odot}^{-1}$ because \scite{brown00} use 4 times the
optical effective radii to extract the X-ray flux, which may actually be
more extended than this).  Thus much of the trend seen by \scite{brown00}
is actually due to the central-dominant group galaxies. In a new study of a
larger sample, \scite{osullivan00} find no trend in L$_X$/L$_B$ with $\rho$
for a sample of 196 early-type galaxies which includes almost all the
\scite{brown00} galaxies.

\begin{figure*}
\psfig{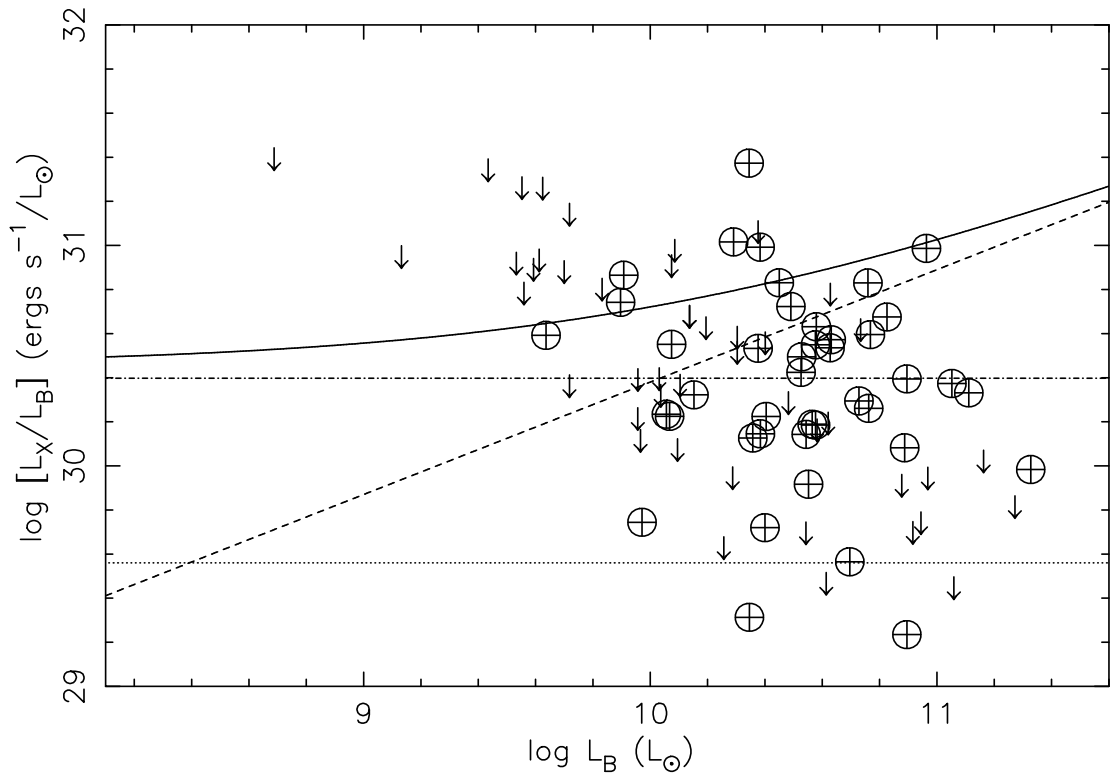}
\caption{\label{fig:lb_lxlb}log L$_X$/L$_B$ vs log L$_B$ for the non
  central galaxies. Dotted line is estimate of discrete source
  contribution, dot-dash line is estimate of energy available from SNIa,
  dashed line is estimate of available energy due to gravitation, and the
  solid line is the sum of the previous three lines.}
\end{figure*}

Our result, then, is that early-type galaxies outside clusters, apart from
central-dominant galaxies in groups, have an apparently universal mean
value of L$_X$/L$_B$, which shows little sign of variation with either
optical luminosity or density of environment. Earlier results to the
contrary appear to be due to the effects of including high luminosity
central-dominant galaxies, or from contamination of L$_X$ values by
intragroup emission.  Once central-dominant galaxies are excluded, the
scatter in L$_X$/L$_B$ is reduced, but as can be seen in
Figure~\ref{fig:lb_lxlb}, it is still substantial, varying over a factor of
20-30 for galaxies of a given optical luminosity.

It is interesting to compare the observed L$_X$/L$_B$ values with the three
lines marked on Figure~\ref{fig:lb_lxlb}. The horizontal dotted line marks
the expected discrete source contribution, discussed in section~5.  Whilst
this line lies close to the lower bound of the data, we note that a number
of galaxies do fall somewhat below it.  \scite{irwin98} also point out that
their derived discrete source contribution shows apparently real
variations, by a factor of at least three, from one galaxy to another.
Attributing this component primarily to low mass X-ray binaries, they argue
that such variations may reflect the abundance of neutron star remnants in
a galaxy, which in turn is sensitive to the initial mass function of its
stars. 

The upper horizontal line in Figure~\ref{fig:lb_lxlb} corresponds to the
energy available from type Ia supernovae. This line is derived using the
\scite{cappellaro99} rate of (0.18$\pm$0.06)$h_{75}^2$ supernovae per
century per $10^{10}$L$_{B\odot}$, where $h_{75}$=H$_0$/75 km s$^{-1}$
Mpc$^{-1}$. Assuming that each supernova releases 10$^{51}$~ergs of energy,
this corresponds to
L$_{SN}$/L$_B$=$2.5\times10^{30}$~erg~s$^{-1}L_{\odot}^{-1}$. This line
lies within the distribution of points, a factor $\sim 2.5$ above the
characteristic mean value of $10^{30}$erg~s$^{-1}$L$_{\odot}^{-1}$ (Figure
6).  What is the source of the additional luminosity in those galaxies
which lie above the SNIa line?  There should be a contribution from the
velocity dispersion of mass-losing stars (the stellar ejecta have an
initial bulk kinetic energy which will be thermalised in the surrounding
interstellar medium) and another from the gravitational work done as the
gas cools and flows towards the centre of the galaxy
\cite{canizares87,brown98}. Both these contributions scale as the square of
the velocity dispersion.  Using the Faber-Jackson relation from
\scite{prugniel96}, and following the analysis of \scite{canizares87},
adopting a King profile galaxy and assuming that the gas flows into the
centre before cooling out ($f=A=1$ in their terminology), we obtain the
`gravitational' line, log$L_{grav}=25.28+1.51 $log$L_B$, marked in
Figure~\ref{fig:lb_lxlb}, which is uncertain by a factor of a few,
depending on the radius at which gas drops out of the cooling flow, and the
mass of the dark galaxy halo.

Adding the discrete source, SNIa and gravitational terms gives an upper
envelope for L$_X$/L$_B$, which is shown in Figure~\ref{fig:lb_lxlb}. Three
of the four galaxies which lie significantly above this line in the plot
are all peculiar. The highest, HCG 15d, is an interacting galaxy which also
a radio source. The HRI emission from this galaxy is dominated by a central
point-like source which is most likely an AGN. The two galaxies above the
line with log(L$_X$/L$_B)\approx 31$ and log L$_B \approx$ 10.3 are a
Seyfert 2 and a starbursting S0 galaxy.

Our conclusion, then, is that all non-central-dominant early-type galaxies
within our sample, apart from a handful which are clearly peculiar,
populate a band in L$_X$/L$_B$ which lies between the discrete source
contribution and the expected luminosity from discrete sources plus a
cooling halo of gas released from galactic stars. This band covers a range
of L$_X$/L$_B$ which changes only weakly with L$_B$ (the lower bound in
Figure~\ref{fig:lb_lxlb} is horizontal, whilst the upper bound rises by
only a factor $\sim 3$ over the range L$_B = 10^9-10^{11}$L$_\odot$), and
where we have reasonable data, our galaxies appear to populate the whole
band. It is therefore not surprising that the mean L$_X$/L$_B$ ratio shows
no significant trend with L$_B$. Larger samples of galaxies would be
required to convincingly resolve any trend associated with the upper
boundary of the band.

The fact that group galaxies (at least in the range L$_B =
10^{10}-11^{10}$L$_\odot$, where we have good coverage) span the full range
from discrete source to full cooling halo lines, indicates that their hot
halos cover a wide range of states. The most X-ray underluminous systems
have either lost all their gas as a result of some recent stripping or star
formation event, or are in a wind phase, in which most of the gas lost by
stars streams out of the galaxy in a fast, low-density wind
\cite{ciotti91}. There is an interesting indication from Figure 5, that
such systems are not represented amongst our sample of genuinely isolated
early-type galaxies, though better statistics are required to be sure. If
confirmed, this observation may relate to the low incidence of tidal
interaction and interaction-induced starburst activity expected in isolated
galaxies, compared to those in groups.  Galaxies with intermediate
L$_X$/L$_B$ values may be in `partial wind' stages \cite{pellegrini98}, and
high resolution X-ray studies with Chandra and XMM-Newton can be used to
search for central cooling flows within such systems. The
non-central-dominant galaxies with the highest values of L$_X$/L$_B$ are
likely to have hot hydrostatic halos with fully developed galactic cooling
flows. However unlike central dominant galaxies, we see no evidence that
these non-central galaxies have excess X-ray luminosity due to accretion of
external gas from the group. Presumably their motion prevents this.

Turning finally to the the central-dominant group galaxies -- these mostly
fall above the upper boundary marked in Figure~\ref{fig:lb_lxlb}. Gas loss
from within the galaxy is unable to explain the high luminosity and
temperature, and the large extent of the X-ray emission in these galaxies,
as pointed out by \scite{brighenti99} and \scite{brown00}. It appears that
additional infalling material is required to adequately reproduce their
observed properties (\pcite{brighenti98},1999\nocite{brighenti99}). The
most likely origin of this infalling material, for dominant group galaxies,
is a group cooling flow, since the X-ray properties of these galaxies
appear to be more closely related to the group than to the galaxy itself
\cite{helsdon01b}.

If this picture is correct, the most X-ray over-luminous early-type
galaxies should be found in the centres of undisturbed bright groups and
clusters.  Other early-type galaxies within galaxy systems should have much
lower values of L$_X$/L$_B$, unless of course they have been the central
galaxy of a group which has recently merged with the present cluster. In
addition, disturbed clusters which show no evidence of any cooling flow
would be expected to contain central galaxies that are less X-ray
overluminous than clusters and groups with cooling flows.

If central group galaxy X-ray properties are more strongly related to the
group than to the galaxy this would account for some of the scatter in the
early-type galaxy L$_X$/L$_B$ relation. It would also explain the
correlation of X-ray luminosity with the relative sizes of the X-ray and
optical emission found by \scite{mathews98}: the most X-ray overluminous
galaxies should be found in the centre of bigger groups.  In addition, the
apparent lack of rotationally enhanced X-ray ellipticity in the cooling
flows of elliptical galaxies \cite{hanlan00}, may be explained if it is a
group, rather than galaxy cooling flow.


\section{Conclusions}
\label{sec:conc}

We have derived the X-ray luminosity for a sample of galaxies in groups
after allowing for the effects of the intragroup emission. This sample is
used to derive the L$_X$:L$_B$ relation for both early- and late-type
galaxies in groups, and these relations are compared to those derived for
other environments.

In general, the X-ray properties of spiral galaxies appear to be
indistinguishable from those in other environments. We find no significant
deviations from the L$_X$:L$_B$ and L$_X$:L$_{FIR}$ relations of spirals
derived from more general samples.

The X-ray properties of early-type galaxies appear to fall into two
distinct categories -- central-dominant group galaxies, and
non-central-dominant galaxies. The non-central group galaxies populate a
band in L$_X$/L$_B$ which shows no discernible trend with optical
luminosity in our data, and is well explained by a combination of emission
from discrete galactic X-ray sources together with a variable contribution
from hot gas released by stars. In contrast the central-dominant group
galaxies are far more X-ray luminous. In the region where the optical
luminosity of the central galaxies overlaps with the non-central sample
their X-ray luminosity is over an order of magnitude greater. It appears
that infall of gas from the intragroup medium is required to account for
these large luminosities.

We have compared the properties of the early-type galaxies in groups with
those in other environments by deriving X-ray luminosities for a sample of
isolated galaxies and also by splitting previous large surveys of
early-type galaxies into group and field subsamples.  Non-central-dominant
galaxies in groups have a L$_X$:L$_B$ relation of slope unity, and the
resulting mean value of L$_X$/L$_B$ appears essentially identical for
isolated and group galaxies. We suggest that steeper L$_X$:L$_B$ slopes
derived in previous work, and apparent dependence of L$_X$/L$_B$ on
surrounding galaxy density, result from the inclusion of central-dominant
group galaxies in galaxy samples.


\section{Acknowledgements}
The authors would like to thank Craig Sarazin for providing pre-publication
information on the spectral characteristics of discrete source populations
in early-type galaxies, and to acknowledge useful discussions with Joel
Bregman and Bill Mathews. We would also like to thank the referee, Jimmy
Irwin, for suggesting several improvements to the paper. The data used in
this work have been obtained from the Leicester database and archive
service (LEDAS). This work made use of the Starlink facilities at
Birmingham and the NASA/IPAC Extragalactic Database (NED). SFH acknowledges
financial support from the University of Birmingham and EJOS acknowledges
the receipt of a PPARC studentship.


\bibliography{../../reffile}

\label{lastpage}

\end{document}